\documentclass[reprint,superscriptaddress]{revtex4-2}
\usepackage{amsmath,amssymb}
\usepackage{mathptmx}
\usepackage{hyperref}
\usepackage{xurl}
\usepackage{graphicx}
\usepackage{makecell}
\usepackage{algorithm}
\usepackage{algpseudocode}
\usepackage{xcolor}
\usepackage{dcolumn}

\newcommand{\ththr}{\ensuremath{\theta_\text{thresh}}}
\newcommand{\Nadj}{\ensuremath{N_\text{adj}}}

\begin{document}

\title{Improved stellarator permanent magnet designs through combined discrete 
       and continuous optimizations}
\thanks{When referencing this work, please cite the published version of 
        record: K.C.~Hammond and A.A.~Kaptanoglu,
       \href{https://doi.org/10.1016/j.cpc.2024.109127}{\textit{Computer 
       Physics Communications} \textbf{299}, 109127 (2024).}}

\author{K.~C.~Hammond}
\email{khammond@pppl.gov}
\affiliation{Princeton Plasma Physics Laboratory, Princeton, NJ 08543, USA}
\author{A.~A.~Kaptanoglu}
\affiliation{University of Maryland, College Park, MD 20742, USA}
\affiliation{Present affiliation: Courant Institute of Mathematical Sciences, 
             New York University, New York, NY 10012, USA}


\begin{abstract}
A common optimization problem in the areas of magnetized plasmas and
fusion energy is the design of magnets to produce a given three-dimensional
magnetic field distribution to high precision. 
When designing arrays of permanent magnets for stellarator plasma confinement,
such problems have tens of thousands of degrees of freedom whose solutions, 
for practical reasons, should be constrained to discrete spaces. 
We perform a direct comparison between two algorithms that have been 
developed previously for this purpose, and demonstrate that composite
procedures that apply both algorithms in sequence can produce substantially
improved results. One approach uses a continuous, quasi-Newton procedure
to optimize the dipole moments of a set of magnets and then projects the
solution onto a discrete space. The second uses an inherently discrete
greedy optimization procedure that has been enhanced and generalized for
this work. The approaches are both applied to design arrays
cubic rare-Earth permanent magnets to confine a quasi-axisymmetric plasma with 
a magnetic field on axis of 0.5~T. The first approach tends to find solutions
with higher field accuracy, whereas the second can find solutions with
substantially (up to 30\%) fewer magnets. When the approaches are combined, 
they can obtain solutions with magnet quantities comparable to the second 
approach while matching the field accuracy of the first.
\end{abstract}

\maketitle

%
%
%
%
%

\section{Introduction}
\label{sec:intro}

The stellarator, a non-axisymmetric toroidal plasma confinement device, is a 
promising concept for a fusion reactor. It has the potential capability to 
operate in a steady state without a requirement for plasma current drive, which
draws a favorable contrast to the axisymmetric tokamak concept. Successful
operation requires good plasma confinement, however, and stellarators must
apply precise, three-dimensional shaping to their magnetic fields to avoid
excessive losses of energy and particles. Accurate field shaping is achieved
in most state-of-the-art stellarators with modular, non-planar coils. Such coils
are complicated to manufacture and assemble \cite{rummel2012a,bosch2013a} and 
can be cost-prohibitive \cite{nielson2010a}. Hence, the development of simpler 
magnets that can be produced at lower cost is a high priority in the 
stellarator research field.

Permanent magnets have recently been proposed \cite{helander2020a} as an 
alternative means of producing three-dimensional stellarator field shaping.
Incorporating permanent magnets could reduce the required compexity of the 
accompanying non-planar coils or, 
in the case of low-field devices, eliminate the need for non-planar coils 
altogether in favor of tokamak-like planar coils. This concept will be explored 
experimentally with MUSE \cite{qian2022a,qian2023a}, a tabletop device recently 
constructed with an array of rare-Earth magnets and a set of circular 
toroidal-field coils.

The permanent magnet stellarator concept gives rise to the question of how
to design an array of permanent magnets that provides the required 
three-dimensional field shaping without being overly complicated to fabricate
and assemble. As with conventional stellarator coil design, this is an
ill-posed problem \cite{landreman2017a} that admits many possible solutions.
Early approaches adapted stellarator coil design methods to specify a toroidal
magnetized volume enclosing the plasma, similar to the ``winding surface'' 
concept often used for modular, non-planar coils \cite{zhu2020a, 
landreman2021a,xu2021a}. More recent approaches relax the requirement of 
toroidal magnet geometry, instead optimizing the individual dipole moments 
of an arbitrary arrangement of magnets \cite{zhu2020b,lu2021a,kaptanoglu2022a}.

The arrays of permanent magnets in these latest designs typically contain a 
large number ($10^3$-$10^5$) of individual magnets. To control the cost and
complexity of fabrication, therefore, it is important for magnet arrays to 
utilize a small number of standardized magnet types rather than having each
magnet be unique in its geometry and/or polarization orientation. This implies
that the design and optimization procedures for the arrays must restrict 
their solutions to discrete spaces, requiring each magnet to have one of only
a few distinct dipole moments. 

Discrete solutions have been achieved to date using a number of different 
approaches. Some approaches employ continuous optimization algorithms 
and subsequently round or project the continuous solutions
onto a discrete subspace \cite{qian2022a,hammond2022a}. Others employ fully
discrete optimization techniques \cite{lu2022a,kaptanoglu2023a}.
Both approaches have produced solutions with good field accuracy for various
quasiaxisymmetric stellarator plasmas. However, as none of these approaches are
guaranteed to find a global optimum, there remains the possibility that 
improved solutions may exist with even more desirable attributes. 

In this paper, we find that hybrid optimization approaches utilizing sequences
of discrete and continuous optimizations can produce solutions that both
exhibit improved field accuracy and require fewer magnets. The two approaches,
when employed on their own, attain solutions with advantages and disadvantages
when applied to a quasiaxisymmetric test case: the ``rounded continuous''
approach achieves greater field accuracy but requires more magnets, whereas
the fully discrete approach can identify solutions utilizing fewer magnets
but with lower field accuracy. When the approaches are combined, in
particular in a sequence of discrete-continuous-discrete, we find that we
are able to obtain solutions with magnet quantities comparable to the low
numbers achieved in the discrete approach while attaining equal or better 
field accuracy metrics than what either approach could achieve on its own.

We will begin by reviewing key aspects of permanent magnet optimization common
to both optimization approaches in Sec.~\ref{sec:common}. We summarize 
the rounded continuous optimization approach in
Sec.~\ref{sec:rc_summary}. We then describe the fully discrete ``greedy''
approach, initially developed in \cite{kaptanoglu2023a} and expanded for this
work, in 
Sec.~\ref{sec:gpmo_enhancements}. In Sec.~\ref{sec:gpmob_results}, results
from optimizations using the enhanced discrete approach are presented and
compared with previous results obtained using the rounded continuous approach.
In Sec.~\ref{sec:combined_approaches}, we present the improved
magnet solutions obtained through combinations of the two approaches.
Finally, in Sec.~\ref{sec:comparison}, we perform a more detailed comparison
of the plasma confinement properties of selected solutions with free-boundary 
equilibrium modeling.

\section{Common aspects of the optimization approaches}
\label{sec:common}

Both the discrete and continuous approaches are applied in this work to solve
the same fundamental problem: given a target stellarator plasma equilibrium
and a set of predefined positions for permanent magnets, what dipole moment
should each magnet have to produce the required field shaping to confine the
plasma? The optimizers seek to choose dipole moments that minimize the 
field error metric $f_B$, an objective function proportional to the square 
integral of the component of the magnetic field normal to the boundary of the 
target plasma:

\begin{equation}
    f_B = \frac{1}{2} \iint_\mathcal{S}
        \left(\mathbf{B}\cdot\mathbf{\hat{n}}\right)^2dA, 
    \label{eqn:fb}
\end{equation}

\noindent Here, $\mathcal{S}$ is the toroidal surface corresponding to the 
boundary of the target plasma equilibrium; $\mathbf{B}$ is the total magnetic 
field, including contributions from the permanent magnets, coils, and plasma 
currents; and $\mathbf{\hat{n}}$ is the unit vector normal to $\mathcal{S}$.
If $f_B = 0$, then the magnetic field is precisely what is needed to
confine the target plasma. In some cases, the optimizer may minimize a 
weighted sum of $f_B$ and other objectives; one example is described in
Sec.~\ref{sec:rc_summary}.

The optimization techniques both take the same key inputs:

\begin{enumerate}
  \item the boundary geometry of the target plasma equilibrium;
  \item contributions to the magnetic field on the target plasma boundary from
        plasma currents and any fixed external magnetic field 
        sources, such as toroidal field coils; 
  \item a list of spatial positions around the plasma where magnets, 
        represented as ideal point dipoles, are eligible to be placed; and
  \item lists of allowable dipole moments for magnets that might be placed in 
        each of the spatial positions.
\end{enumerate}

\noindent We note that, while the dipole moments in the solution produced by 
each optimization procedure must match allowable moments in their respective 
lists as described in Item 4, they may deviate from the lists during 
intermediate steps of the procedure. This is indeed the case for the rounded 
continuous approach in Sec.~\ref{sec:rc_summary}.

\begin{figure}
  \begin{center}
  \includegraphics[width=0.5\textwidth]{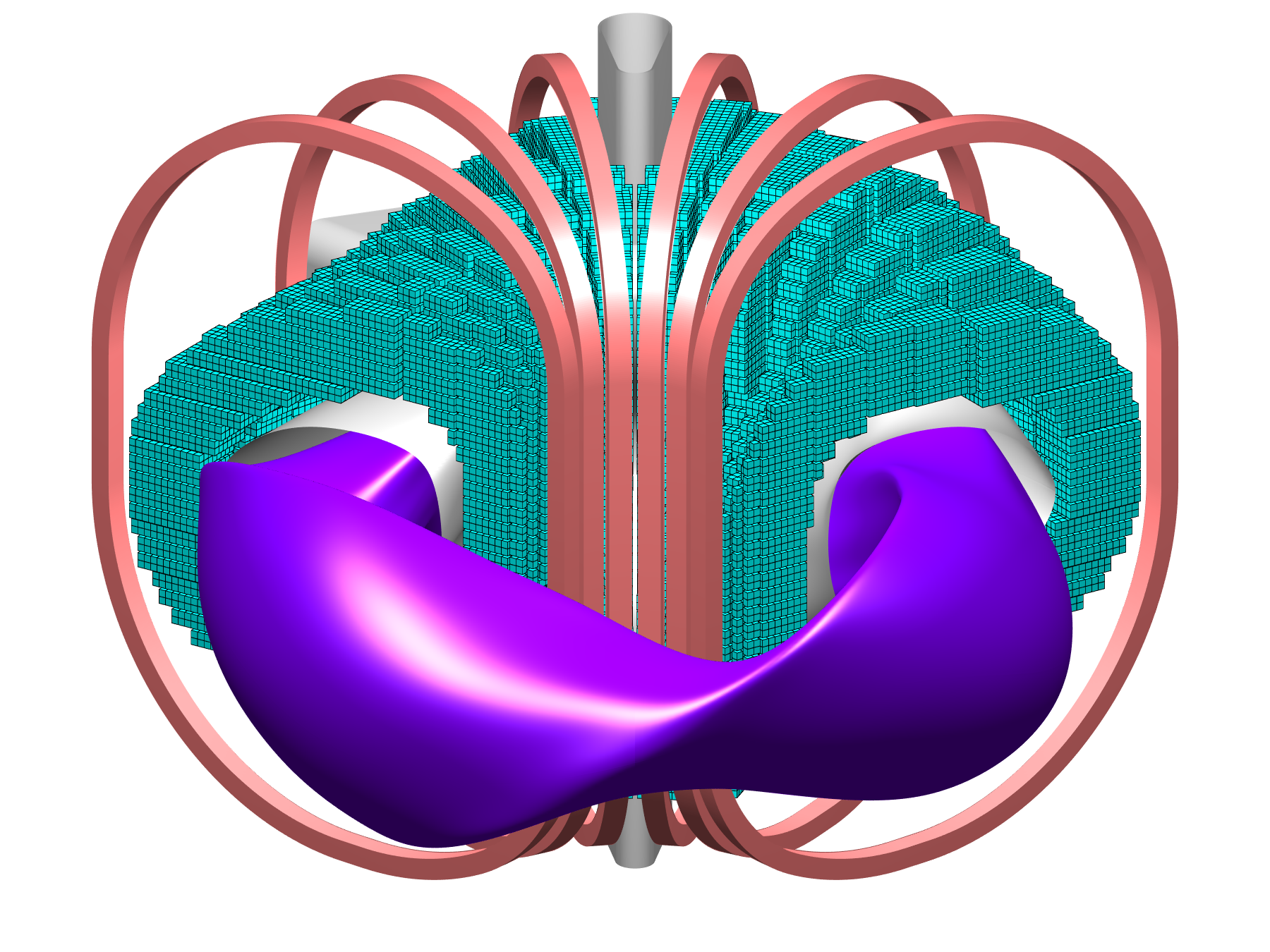}
  \caption{Renderings of the plasma equilibrium (purple), toroidal field coils
           (pink) and the arrangement of permanent magnets (cyan) used for the
           optimizations in this paper. The plasma vessel and ports are also 
           shown in gray. Coils and magnets are shown for one field period 
           only.}
  \label{fig:setup} 
  \end{center}
\end{figure}

Items 1-3 are illustrated in Fig.~\ref{fig:setup}. In this work,
the target plasma equilibrium is similar to that of NCSX
\cite{zarnstorff2001a,nelson2003a} but with the magnetic field on axis scaled
down to 0.5~T. The configuration has a major radius of 1.44~m, a minor radius
of 0.32~m, and a volume-averaged plasma $\beta$ of 4.1\%.
It exhibits stellarator symmetry \cite{dewar1998a}
and has three field periods, each of which consists of two equivalent 
half-periods. A set of eighteen planar toroidal-field (TF) coils
with fixed currents are assumed to supply the toroidal magnetic field.
The arrangement of candidate magnet positions was generated by the 
\textsc{Magpie} code \cite{hammond2020a}. The arrangement contains a total of
349,548 possible magnet positions distributed around the torus, or 
58,258 per half-period. The magnets are assumed to each
be cubes with a side length of 3~cm, and are grouped into 48 sectors (8 per
half-period), each of which consist of vertically stacked blocks. Further 
details about this specific magnet arrangement are given in \cite{hammond2022a}.

\begin{figure}
  \begin{center}
  \includegraphics[width=0.25\textwidth]{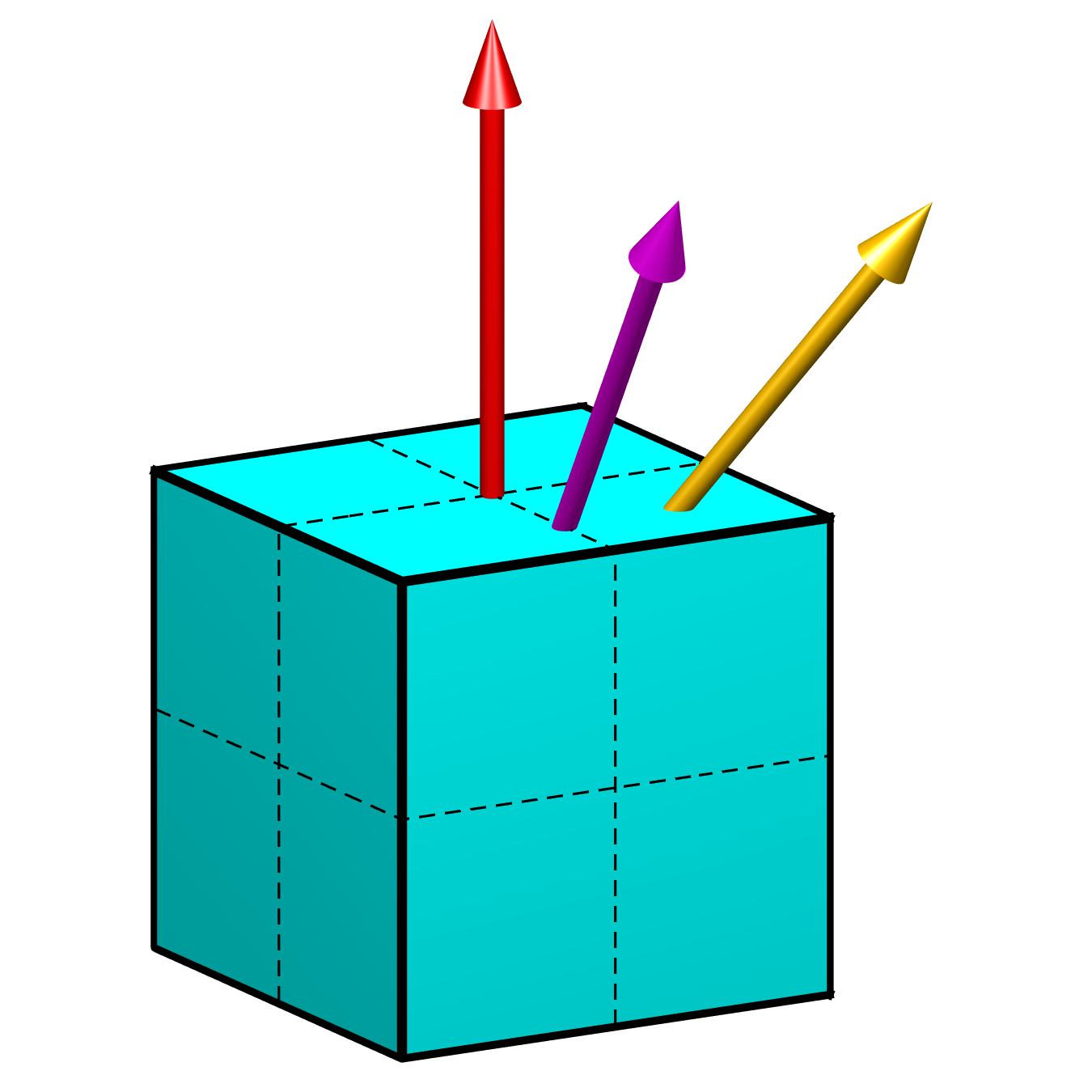}
  \caption{Schematic of the three polarization types allowed for the cubic
           magnets appearing in the solutions studied in this paper, including
           face (red), face-edge (magenta), and face-corner (yellow). 
           Reprinted from Ref.~\cite{zhu2022a} with the permission of AIP
           Publishing.}
  \label{fig:types}
  \end{center}
\end{figure}

Item 4 is illustrated schematically in Fig.~\ref{fig:types}. The allowable
dipole moments for each magnet in the arrangement are defined such that the
solution may contain at most three unique magnet types: polarization 
perpendicular to two faces (\textit{face} type), polarization in a plane
parallel to two faces (\textit{face-edge} type), and polarization in a plane
that contains two opposite edges (\textit{face-corner} type). With the geometric
orientation of each magnet fixed, for each magnet there are six possible
dipole moments of the face type, twenty-four of the face-edge type, and
twenty-four of the face-corner type; thus, there are a total of fifty-four
allowable dipole moments for each magnet in the arrangement. Further details 
on the polarization types and how they were chosen are given
in \cite{hammond2022a}.  

The optimization approaches studied in this paper all calculate the field
from each permanent magnet by approximating it as an ideal magnetic dipole
located at the magnet's centroid. The dipole moment vector may be written as

\begin{equation}
  \mathbf{m} = \rho M V \mathbf{\hat{v}},
  \label{eqn:dipole_moment}
\end{equation}

\noindent where $\rho$ is a scaling constant between 0 and 1, 
$M$~=~1.10~MA/m is the
magnetization of a typical rare-Earth magnet (corresponding to a remanent
magnetic field $B_r$ of 1.38~T), $V$ is the magnet volume, and 
$\mathbf{\hat{v}}$ is a unit vector in the direction of the magnet's
polarization. In the solution space permitted for both optimization approaches, 
$\rho$ is constrained to either be zero (in which case the magnet does not 
exist) or 1. However, during the course of the optimization, one approach 
allows $\rho$ to vary continuously, as described in more detail in 
Sec.~\ref{sec:rc_summary}.

The dipole approximation described above neglects the effects of
finite magnet size and the difference of the magnets' permeability from the 
vacuum permeability $\mu_0$. However, higher-fidelity finite-element analyses 
have indicated that the discrepancies between the magnetic field on the 
plasma boundary calculated with and without the assumption of ideal dipoles
are small compared the anticipated sources of field error due to fabrication
imperfections and assembly tolerances \cite{zhu2022a}, which in turn can be 
corrected if necessary by an auxiliary set of magnets \cite{rutkowski2023a}.

\section{Summary of the rounded continuous optimization approach}
\label{sec:rc_summary}

The optimization procedure for permanent magnets that we refer to in this paper
as the ``rounded continuous'' approach is described in detail in 
\cite{hammond2022a}, and we summarize it in this section for completeness.
The rounded continuous approach involves two successive continuous 
optimizations of the dipole moments of each magnet in the arrangement, 
followed by a projection of the continuous solution onto the discrete solution 
space.

The continuous optimizations utilize a quasi-Newton algorithm \cite{zhu1997a}
implemented in the \textsc{Famus} code \cite{zhu2020b}. The algorithm adjusts 
up to three independent paramenters for each magnet in order to minimize an 
objective function. These parameters include the scaling factor $\rho$ for the 
dipole moment magnitude (Eq.~\ref{eqn:dipole_moment}), the azimuthal 
orientation angle 
$\phi$, and the polar orientation angle $\theta$. The scaling factor $\rho$ is 
constrained to have an absolute value between 0 and 1 inclusive. Before the 
optimization, initial
values must be chosen for each parameter. The solution is known to vary for
different initializations, indicating a solution space with many local minima
\cite{hammond2020a}. In absense of prior knowledge, we typically initialize 
$\rho$ as 0 for each magnet and set the angular parameters $\phi$ and $\theta$
to be oriented approximately perpendicular to a nearby point on the vacuum
vessel enclosing the plasma. 

The two successive continuous optimizations performed in the rounded continuous
approach differ in the choice of objective function. In the first stage, the
objective function is equal the field error metric 
$f_B$ (Eq.~\ref{eqn:fb}). The result of this optimization may have
substantially improved field accuracy; however, due to the continuous nature
of the procedure the dipole moments are likely all unique in strength $\rho$
and orientation angles $\phi$ and $\theta$. Thus, a second optimization 
is performed with a weighted sum of two objectives: 
$f_B~+~\lambda f_\rho$, where $\lambda$ is a weighting factor and

\begin{equation}
  \label{eqn:frho}
  f_\rho = \sum_i^N \left(|\rho_i|(1 - |\rho_i|)\right)^2
\end{equation}

As indicated in Eq.~\ref{eqn:frho}, $f_\rho$ penalizes intermediate
values of the magnitude scaling factor $\rho$, favoring solutions in which
$\rho$ is either 0 or 1. The weighting factor $\lambda$ is chosen such that
the second optimization maintains a low value of $f_B$ while reducing
$f_\rho$ and thereby (hopefully) achieving an accurate solution in which
all magnets have either zero strength or full strength. 

While the elimination of magnets with arbitrary, intermediate strengths greatly 
simplifies fabrication, the solution at the second optimization stage still
exhibits another complicating feature. Specifically, since the orientation
angles $\phi$ and $\theta$ for each dipole moment were allowed to vary 
continuously without constraints or penalization, each magnet likely has a 
unique polarization orientation. The rounded continuous approach thus entails
one final step to further simplify the fabrication requirements: the 
orientation of each (nonzero) dipole moment in the solution is rounded to
the nearest of a set of allowable orientation vectors. Assuming the
allowable orientation vectors correspond to a limited set of polarization
types such as the one illustrated in Fig.~\ref{fig:types}, this has the
effect of projecting the solution into a discrete space with a low number
of unique magnet types.

\begin{figure}
  \begin{center}
  \includegraphics[width=0.49\textwidth]{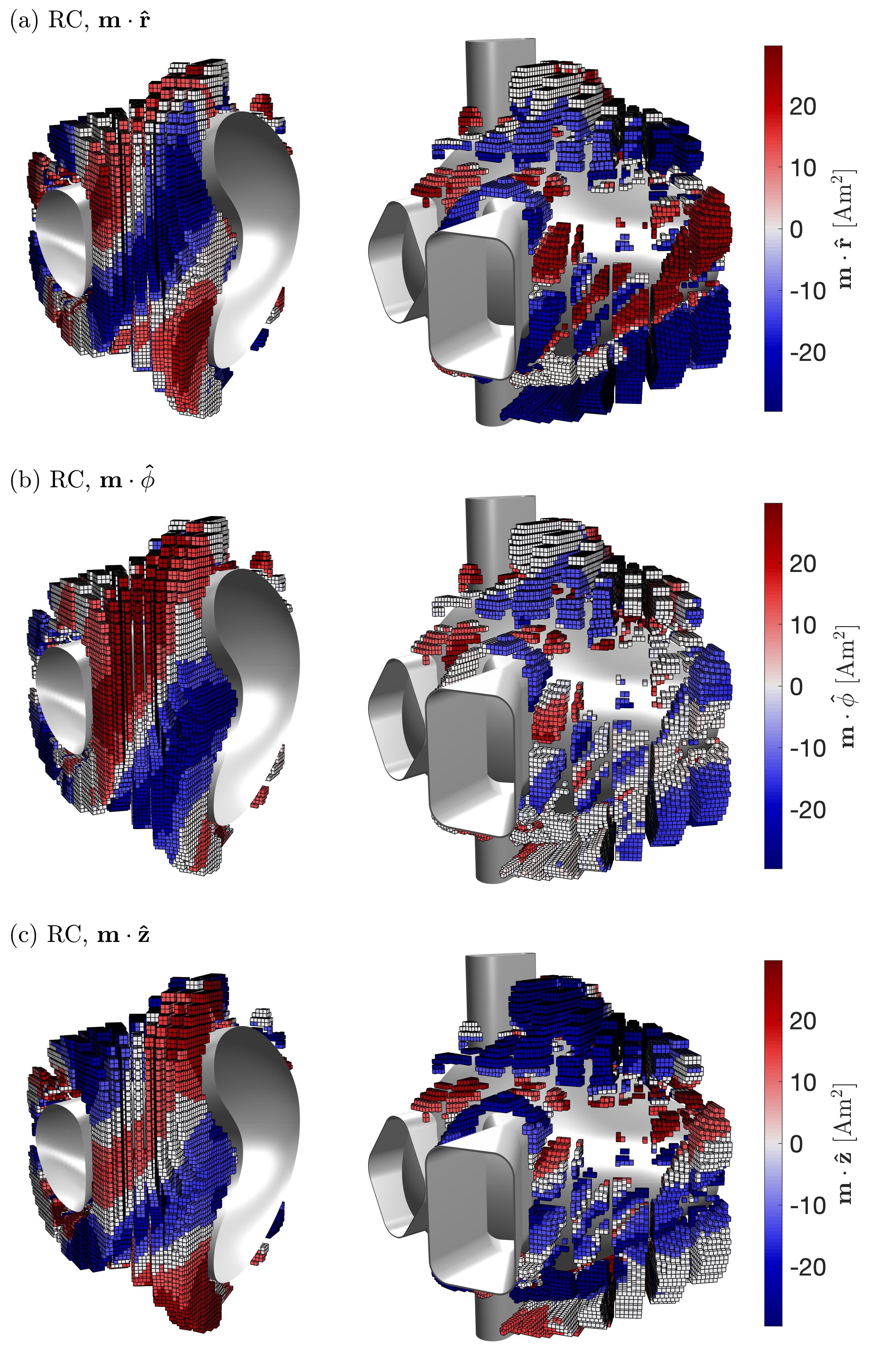}
  \caption{Renderings of the magnets in the solution produced by the rounded
           continuous (RC) optimization approach in Sec.~\ref{sec:rc_summary},
           color-coded according to different cylindrical components of their
           dipole moments. The left and right columns show views of one 
           half-period from the inboard and outboard sides, respectively.
           The outboard view also contains renderings of access ports where
           magnets could not be placed.
           (a) radial $r$ component, 
           (b) toroidal $\phi$ component,
           (c) vertical $z$ component.}
  \label{fig:vectors_rc}
  \end{center}
\end{figure}

This optimization approach was developed and applied in \cite{hammond2022a}
to design an array of magnets for the NCSX-like plasma configuration shown
in Fig.~\ref{fig:setup}. To uphold the configuration's stellarator symmetry,
the magnets were constrained to be uniform across half periods; hence,
the procedure was effectively optimizing within an arrangement of 58,258
possible magnet positions (i.e. the number of possible positions within one
half-period). The solution retained 35,436 magnets per half-period, 
corresponding to 212,616 magnets for the full torus with a total magnetized
volume of 5.74~m$^3$. 

Renderings of this solution, showing the magnet positions, orientations, 
and dipole moments for one half-period of the torus, are shown in 
Fig.~\ref{fig:vectors_rc}. Note that the arrangement of magnet positions
leaves space for multiple access ports on the top, bottom, and outboard
sides (right column), illustrating the flexibility of the permanent magnet
concept and associated optimization methods for accommodating arbitrary
spatial requirements for other device components. In addition, many of the
available magnet positions on the outboard side are left unused, as can be
seen in the gaps in the magnets through which the vacuum vessel is visible.
By contrast, the magnet positions on the inboard side of the torus
(left column) are nearly fully occupied.

For this solution, the two continuous optimizations in \textsc{Famus} each 
determined $\rho$, $\theta$, and $\phi$ for dipoles at each of the 58,258 
positions in the input arrangement, constituting a total of 174,774 free 
parameters. The field error metric $f_B$ was estimated based on evaluations of
$\mathbf{B}\cdot\mathbf{\hat{n}}$ at 8,192 test points on the target plasma
boundary. The optimizations were each run for 200 quasi-Newton iterations.
Using 64 cores on a 2.9 GHz Intel Cascade Lake processor node 
on the Stellar cluster at Princeton University \cite{stellar_cluster}, the 
first-stage \textsc{Famus} optimization took 
53 minutes and the second-stage optimization took 55 minutes. The final 
discretization step took about 2 seconds on a single core. Thus, the full
rounded continuous optimization procedure had a wall clock time of 108 minutes, 
corresponding to 115 CPU-hours.

The solution had a field error metric of $f_B=1.20\times10^{-4}$~T$^2$m$^2$.
Prior to the final discretization step (i.e. after the second continuous
optimization), the error metric was $3.76\times10^{-5}$~T$^2$m$^2$. The increase
in $f_B$ following the discretization is to be expected, as the discretization 
modified the dipole moments with no requirement to reduce or maintain $f_B$.
Nevertheless,
free-boundary equilibrium modeling using the magnetic field produced by the
discretized solution indicated that its field accuracy was 
sufficient for confining the target plasma equilibrium. This solution will be 
designated as ``RC'' when comparing with other solutions in this paper.

\section{Enhancements to the greedy permanent magnet optimization approach}
\label{sec:gpmo_enhancements}

The rounded continuous optimization approach discussed in the previous 
section performs the optimization in a continuous solution space and then
projects the continuous solution onto a discrete space. By contrast, the 
greedy permanent magnet optimization (GPMO) algorithm works exclusively
in the discrete solution space defined by the allowable polarization vectors
assigned to each magnet in the arrangement. In this section, we summarize
the GPMO algorithm introduced in \cite{kaptanoglu2023a} and describe some
generalizations we have implemented to enable the results obtained in this
paper.
We also note that the techniques described here follow an approach conceptually 
similar to the ``two-step'' permanent magnet optimization algorithm developed 
earlier by Lu et al.~\cite{lu2021a,lu2022a}.

The basic GPMO algorithm seeks to minimize the objective function $f_B$ by
adding magnets one by one. In each iteration, the algorithm cycles through
the available positions in the arrangement and calculates the effect on 
$f_B$ of adding a magnet at each position with every allowable dipole moment
vector associated with the respective position. It then adds the magnet
with the position and dipole moment that results in the greatest reduction 
of $f_B$. This process repeats until there are no more available positions
for magnets or until the total magnet quantity reaches a user-defined maximum.
For the stellarator-symmetric example explored in this paper, iterations really
add six magnets at a time, as placing a magnet with a given position and 
moment vector in one half-period requires magnets with equivalent positions
and orientations to be placed in the remaining five half-periods to uphold the
symmetry.

This conceptually simple approach tends to be very effective at reducing 
$f_B$ during early stages of the optimization when few magnets are present in
the array and $f_B$ is high. However, since the algorithm adds each magnet
individually and does not consider collective contributions of groups of
magnets to the overall field shaping, the solution can become suboptimal
as more magnets are added. A typical symptom of this suboptimality is the
presence of \textit{conflicting pairs} of magnets; i.e. nearby magnets that 
have dipole moments pointing in opposite (or near-opposite) directions. Such
conflicting pairs make minimal contributions to the field shaping at points
on the plasma boundary that are far away compared to the distance between
the two magnets. This makes inefficient use of the magnet mass in the array,
and limits the potential for accurate field shaping.

An effective remedy for conflicting pairs is a procedure called 
\textit{backtracking}. The backtracking procedure searches through the
magnets within the solution for conflicting pairs and removes them. Positions
in the arrangement formerly occupied by conflicting magnets become eligible
for new magnets to be placed during subsequent GPMO iterations. 
Presumably, these replacement magnets would have orientations that are
more suitable in the context of the other neighboring magnets that have 
already been placed.
Performing the backtracking procedure periodically after a certain number of
GPMO iterations has been shown to lead to solutions with substantially lower
$f_B$. We will hereafter refer to optimizations employing the GPMO algorithm 
with occasional backtracking as ``GPMOb''.

The behavior of the backtracking procedure is regulated through two key
parameters.  The first parameter, $\ththr$, is the minimum angle
between two polarization vectors for them to be considered conflicting.
If $\ththr=180^\circ$, for example, then only 
magnets with directly opposing dipole moment vectors are eligible for removal.
Reducing $\ththr$ would tend to increase the number of nearby pairs of magnets
eligible for removal. This would make the optimization more stringent, in 
the sense that the backtracking procedure will have a lower tolerance for 
differences in
the dipole moments among nearby magnets and therefore remove more magnets 
overall. The second parameter, $\Nadj$, is the number of nearest neighboring 
positions of a given magnet in which to check for magnets with a conflicting 
polarizations. In general, the larger $\Nadj$, the further apart conflicting
magnets can sit within the arrangement and still be eligible for removal during 
backtracking. Thus, a higher value of $\Nadj$ leads to a more stringent
optimization with more conflicting pairs identified and removed.
In subsequent discussions, we will sometimes use the notation 
GPMOb$_{\ththr,\Nadj}$ to refer to solutions obtained using the GPMOb
algorithm with given values of $\ththr$ and $\Nadj$.
 

The GPMO and GPMOb algorithms were recently implemented \cite{kaptanoglu2023a} 
in the open-source \textsc{Simsopt} software base \cite{simsopt}. 
For the work in this paper, the implementation was generalized to allow
for the user to specify an arbitrary, customizable set of allowed dipole 
moment vectors for each position in a magnet arrangement. This enables, 
for example, designs that utilize the three types of magnet polarizations
shown in Fig.~\ref{fig:types}. In addition, we have enabled an arbitrary 
choice of $\ththr$ for the backtracking procedure. As shown in the 
following sections, this flexibility greatly expands the variety of solutions
that can be found through this method. Finally, the implentation now supports
arbitrary initial guesses for the solution (previously, all magnets in the 
arrangement were initialized to have dipole moment of zero). This enables
the usage of GPMOb for refining solutions obtained through other methods,
such as the rounded continuous procedure.

\section{GPMOb results}
\label{sec:gpmob_results}

\begin{figure}
  \begin{center}
  \includegraphics[width=0.48\textwidth]{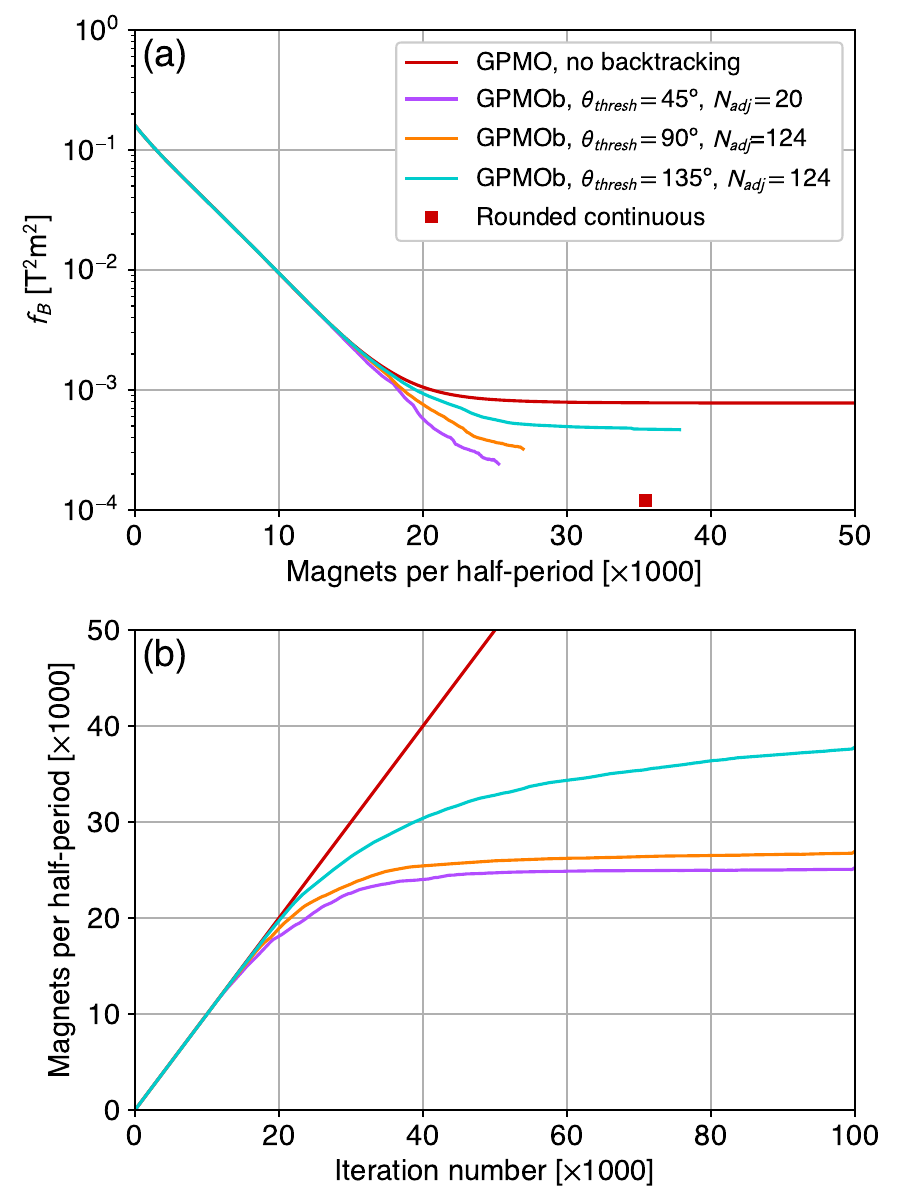}
  \caption{Iteration data from four example greedy permanent magnet 
           optimizations admitting solutions with the three polarization types
           shown in Fig.~\ref{fig:types}. Note that only one in about 400 
           iterations are plotted for each greedy optimization.
             (a) Values of the field-error objective function $f_B$ shown as a 
                 function of the number of magnets per half-period at selected 
                 iterations. For reference, a data point is also
                 shown for the rounded continuous solution described in 
                 Sec.~\ref{sec:rc_summary} (red square).
             (b) Data from the same greedy optimizations, this time showing
                 magnet quantity versus iteration number.}
  \label{fig:iterations_gpmob_ftri}
  \end{center}
\end{figure}

Data from some example greedy permanent magnet optimizations for the target
quasi-axisymmetric plasma in Fig.~\ref{fig:setup} are shown in
Fig.~\ref{fig:iterations_gpmob_ftri}. The optimizations placed magnets in the
arrangement from Fig.~\ref{fig:setup}, which was initialized to be empty in
each case. The optimizations differed in only in the values of 
$\ththr$ and $\Nadj$ used during the backtracking procedures,
which were conducted after every 200 greedy iterations. In addition, one
optimization was performed without backtracking (red curves). In every 
case, the solutions admitted magnets of the three polarization types (with
a total of 54 possible dipole moments) shown
in Fig.~\ref{fig:types}.

Each point on the curves in Fig.~\ref{fig:iterations_gpmob_ftri} represents one
GPMO iteration. Fig.~\ref{fig:iterations_gpmob_ftri}a shows the field error 
metric $f_B$ in relation to the number of magnets per half-period for each of 
these iterations, while Fig.~\ref{fig:iterations_gpmob_ftri}b shows the number 
of magnets plotted against the iteration number. When no backtracking is 
performed, the number of magnets per half-period in the solution is simply 
equal to the number of iterations, as can be seen in the red curve in 
Fig.~\ref{fig:iterations_gpmob_ftri}b. However, when backtracking is performed,
magnets are periodically removed from the solution and therefore the number
of magnets per half-period will be less than the iteration number. While these
curves have a smooth appearance in Fig.~\ref{fig:iterations_gpmob_ftri}b, note 
that they contain data from only roughly every 400th iteration. If every 
iteration had been plotted, a discontinuous drop in magnet quantity would 
appear after each backtracking procedure (occurring every 200th iteration).

Fig.~\ref{fig:iterations_gpmob_ftri} highlights the effects that the
backtracking parameters $\ththr$ and $\Nadj$ have on the solutions
that the GPMOb algorithm ultimately achieves. 
The solutions are all essentially
the same for about the first 15,000 iterations, but diverge from one another
thereafter.
Most strikingly, the ultimate magnet quantity tends to be lower
for lower values of $\ththr$ and for higher values of $\Nadj$.
This is to be expected, as both of those trends reflect more expansive
definitions of conflicting magnet pairs: for lower $\ththr$, magnets
with smaller differences in polarization are considered conflicting, and
for higher $\Nadj$, magnets that are further apart can be considered
conflicting. In other words, decreasing $\ththr$ and increasing
$\Nadj$ both increase the number of magnets subject to removal during
the backtracking procedure. 

Note that if the backtracking parameters are sufficiently stringent, the 
solution will converge to a terminal magnet quantity lower than the amount 
otherwise permitted by the optimization parameters. This indicates that the 
solution has reached a point in which every additional magnet added in 
subsequent greedy iterations would conflict with magnets that are already in 
the solution. As can be seen in Fig.~\ref{fig:iterations_gpmob_ftri}b, this 
terminal quantity is roughly 25,000 magnets per half-period for 
$(\ththr=45^\circ,~\Nadj=20)$ and 27,000 magnets per half-period for 
$(\ththr=90^\circ,~\Nadj=124)$.

Another key difference among the greedy optimizations is in the level of
field accuracy attained in the solutions. Without backtracking, the field
error objective $f_B$ levels off at $7.79\times10^{-4}$~T$^2$m$^2$, whereas 
the solution with $(\ththr=45^\circ,~\Nadj=20)$ reaches 
$2.42\times10^{-4}$~T$^2$m$^2$,
an improvement by a factor of more than 3. The dependence of the attainable
$f_B$ on $\ththr$ and $\Nadj$ is complex. To find the best values for $\ththr$
and $\Nadj$, we performed a more extensive scan of GPMOb optimizations with
wider ranges of values for each parameter. The results are discussed in 
more detail in Appendix \ref{sec:btparams}. 

From this scan, the best field accuracy for our target plasma was obtained using
$\ththr=45^\circ$ and $\Nadj=20$, with $f_B = 2.42\times10^{-4}$~T$^2$m$^2$. 
However, the optimal choice of $\Nadj$ and 
$\ththr$ is in general contingent on the target plasma, the arrangement of 
magnet positions, and the allowable polarization vectors at each position.
Thus, when designing a magnet array for a new stellarator plasma
with different allowable polarization vectors for the magnets, it will be
important to scan through the space of $\Nadj$ and $\ththr$ to find the values
that yield the best solutions.

\begin{figure}
  \begin{center}
  \includegraphics[width=0.49\textwidth]{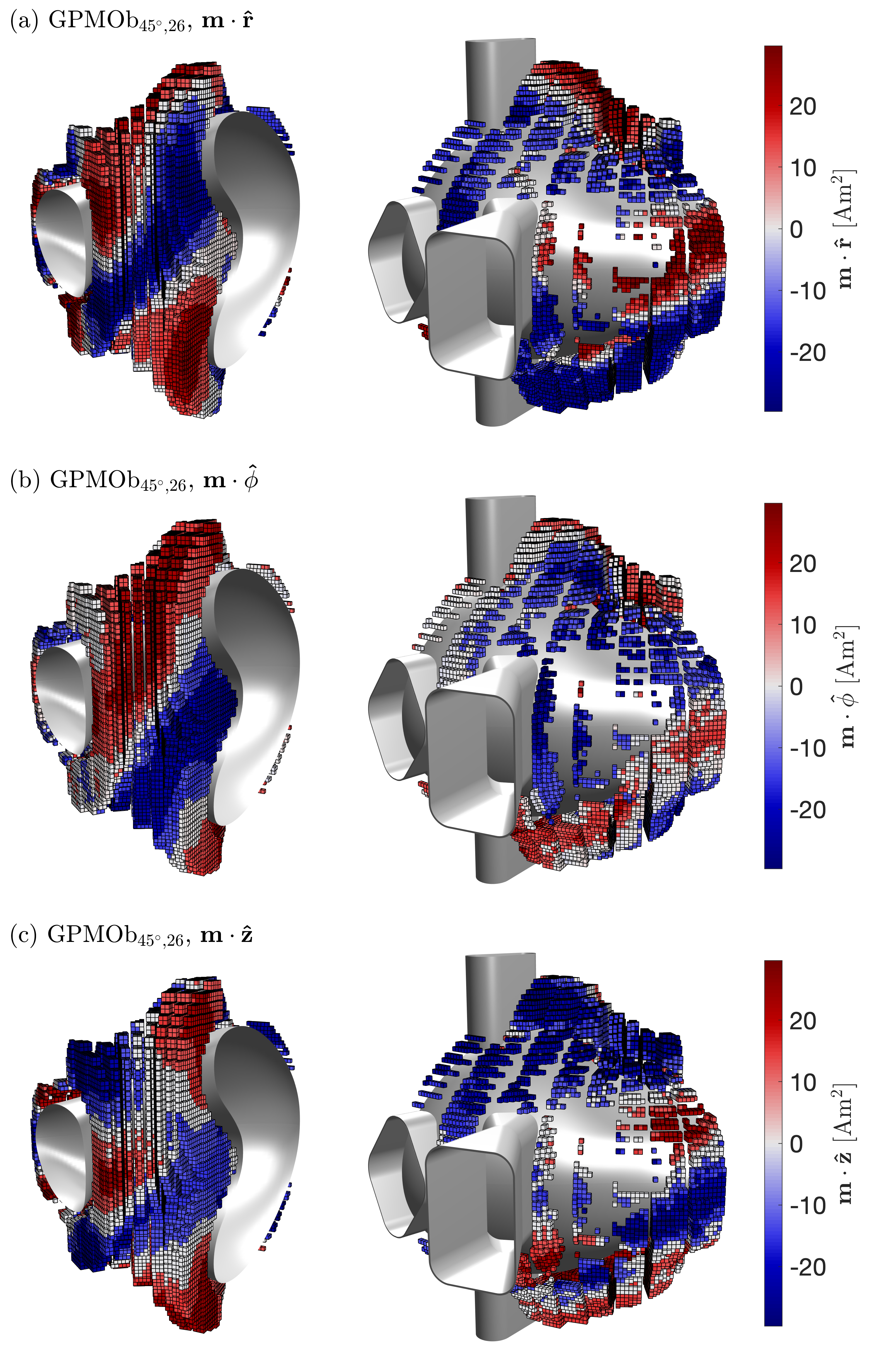}
  \caption{Renderings of the magnets in the solution produced by the greedy
           permanent magnet optimization approach with backtracking (GPMOb)
           using parameters $\ththr=45^\circ$ and $\Nadj=26$, as described
           in Sec.~\ref{sec:gpmob_results}.
           (a) radial $r$ component, 
           (b) toroidal $\phi$ component,
           (c) vertical $z$ component.}
  \label{fig:vectors_gpmob_45_26}
  \end{center}
\end{figure}

Renderings of the solution with $\ththr=45^\circ$ and $\Nadj=26$ are shown
in Fig.~\ref{fig:vectors_gpmob_45_26}. Remarkably, the spatial distributions
of each dipole moment vector component are qualitavely very similar to those
of the RC solution (Fig.~\ref{fig:vectors_rc}), particularly on the inboard
side. The most notable difference is that the GPMOb solution has far fewer
magnets on the ouboard side, particularly near the top and bottom.

The GPMOb optimizations described in this section were each run for 
100,000 iterations and used 4,096 test points on the plasma boundary to
evaluate $f_B$. For the magnet arrangement with 58,258 positions and
54 different allowable polarization vectors at each position, an optimization
initialized to an empty magnet array and using parameters $\ththr=45^\circ$
and $\Nadj=26$ took 4 hours and 20 minutes to run using 64 CPUs 
on a 2.9 GHz Intel Cascade Lake processor node \cite{stellar_cluster}, 
corresponding to 264-CPU hours. 
For a more direct comparison with the time tests for the RC procedure in 
Sec.~\ref{sec:rc_summary}, we re-ran the optimization with 8,192 test points on 
the plasma boundary. This had a wall clock time of 8 hours and 20 minutes, 
corresponding to 533 CPU-hours. 

The wall clock time also exhibited a strong
dependence on the choice of backtracking parameters. For example, a GPMOb
run with $\ththr=45^\circ$, $\Nadj=2$ took roughly 40\% less time to complete
than a run with $\ththr=45^\circ$, $\Nadj=26$. The faster processing time
for the former case is due to the fact that the lower setting of $\Nadj$
causes fewer magnets to be removed during each backtracking procedure. As a 
result, fewer empty spaces remain to be checked in subsequent greedy iterations.

As a quantitative reference for comparison to the solutions obtained with the 
GPMOb approach, the $f_B$ and magnet quantity for the RC solution described in 
Sec.~\ref{sec:rc_summary} is plotted as the red square in 
Fig.~\ref{fig:iterations_gpmob_ftri}. While some of the GPMOb solutions 
require substantially fewer magnets than the RC solution, they do not quite
match its field accuracy: the lowest $f_B$ value obtained with GPMOb was
$2.42\times10^{-4}~$T$^2$m$^2$, whereas the RC solution had
$f_B=1.20\times10^{-4}~$T$^2$m$^2$. Hence, the solutions obtained with the
two different methods exhibit a trade-off. The GPMOb solution would be
cheaper to construct by virtue of having fewer magnets, while the 
RC solution would exhibit better plasma confinement due to its higher field
accuracy. In the next section, we will explore how performing hybrid
optimizations that make use of both approaches can attain solutions with
both fewer magnets and higher field accuracy.


\section{Improved solutions with combined approaches}
\label{sec:combined_approaches}

The differing advantages and disadvantages of solutions obtained by the 
RC and GPMOb approaches motivated a study of whether one approach can be used
to improve upon a solution obtained by the other. In this 
section, we present the results of this study. In Sec.~\ref{sec:rc-gpmob},
it is shown that applying the GPMOb algorithm to a solution first obtained
with the RC approach can both improve field accuracy and reduce the number
of magnets in the solution. Then, in Sec.~\ref{sec:gpmob-rc-gpmob}, we show
that applying the RC algorithm to a GPMOb solution, and then fine-tuning that
solution with another round of GPMOb, can produce a solution that maintains
the low magnet count of the initial GPMOb solution while exhibiting 
substantially lower objective function values.

\subsection{RC-GPMOb}
\label{sec:rc-gpmob}

\begin{figure}
  \begin{center}
  \includegraphics[width=0.48\textwidth]{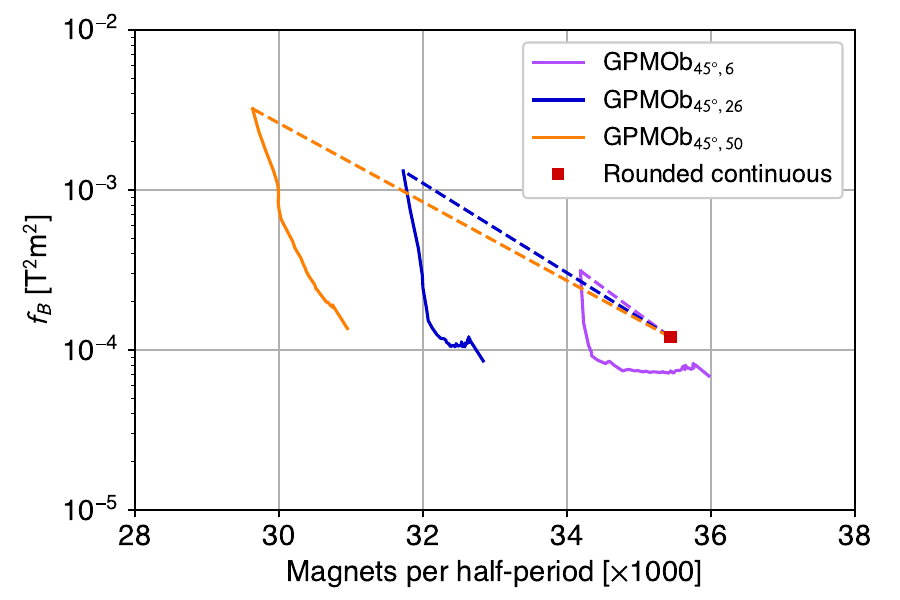}
  \caption{Iteration data from three exaple GPMOb optimizations initialized
           with the RC solution from Sec.~\ref{sec:rc_summary}. Dashed lines
           represent the change in $f_B$ and magnet quantity after applying
           backtracking to the RC solution. 
           For reference, a data point is also shown for the 
           rounded continuous solution (red square).}
  \label{fig:iterations_rc_gpmob}
  \end{center}
\end{figure}

The GPMOb algorithm was applied with various $(\ththr,\Nadj)$ values to the 
solution obtained with the RC approach from Sec.~\ref{sec:rc_summary}.
In each case, the RC solution was input to the GPMOb algorithm as an initial
guess. 
Evolution of $f_B$ and magnet quantity for some of these GPMOb optimizations
is plotted in Fig.~\ref{fig:iterations_rc_gpmob}. The first backtracking
procedure performed after initialization checks the full RC solution for
conflicting magnet pairs. The resulting removal of magnets leads to a 
drop in magnet quantity and a rise in $f_B$ relative to the RC solution.
As indicated by the dashed
lines in the figure, these shifts vary depending on $\ththr$ and $\Nadj$.
Following the loss of magnets in the first backtracking procedure, the GPMOb 
algorithm is typically
able to recover the loss of field accuracy, in some cases even outperforming
the initial $f_B$ value from the RC solution. In addition, while the solutions
typically regain some of the magnets lost in the initial backtracking 
procedure, the GPMOb optimization is in many cases able to nearly match
or even exceed the field accuracy of the RC solution with fewer magnets
overall.

We remark that the optimizations in Fig.~\ref{fig:iterations_rc_gpmob} 
exhibit a sudden drop of $f_B$ and increase in quantity on the final plotted
iteration. This occurs because the solution emerging from the final iteration
is not subjected to backtracking, which would have reduced the quantity and
increased $f_B$. Data plotted from all preceding iterations were taken from
iterations occurring directly after a round of backtracking.

\begin{figure}
  \begin{center}
  \includegraphics[width=0.48\textwidth]{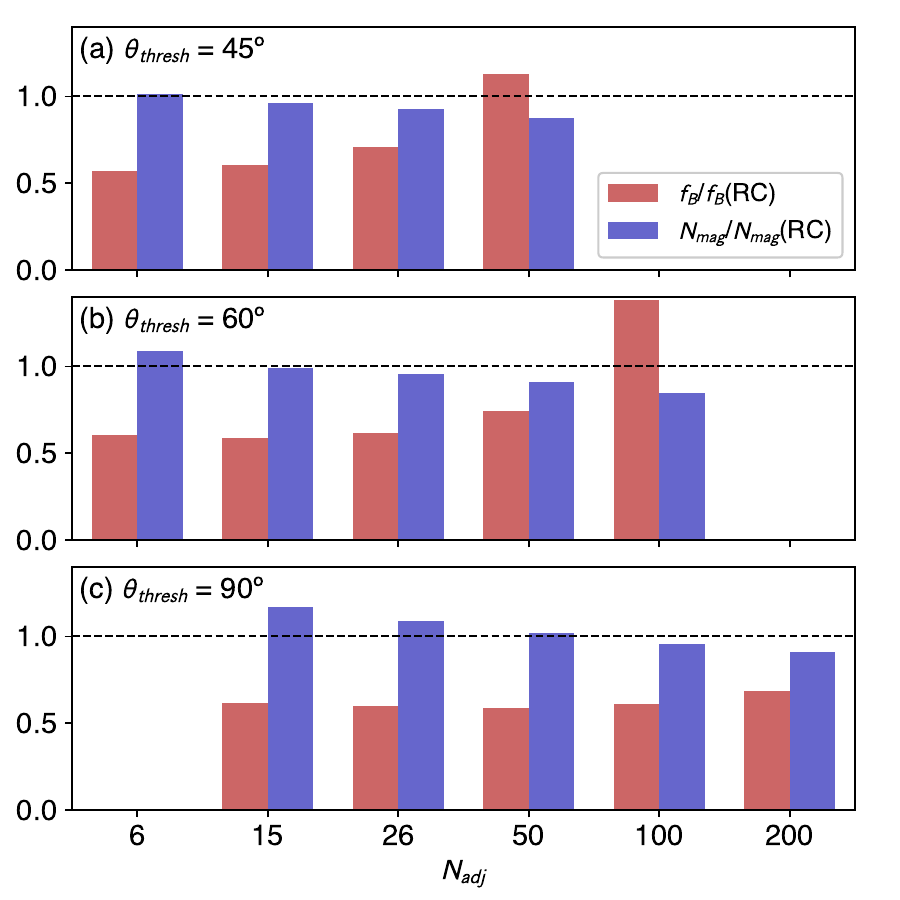}
  \caption{Performance of GPMOb optimizations with different $\ththr$ and
           $\Nadj$, all initialized to the RC solution as described in
           Sec.~\ref{sec:rc-gpmob}. For each GPMOb solution, values of the error
           field objective $f_B$ and magnet quantity $N_\text{mag}$ are shown
           relative to the corresponding values for the RC solution:
           $f_B(RC) = 1.20\times10^{-4}$~T$^2$m$^2$, 
           $N_\text{mag}(RC)=$~35,436 magnets/half-period.}
  \label{fig:fb_nm_rc_gpmob}
  \end{center}
\end{figure}

The performance of the RC-initialized GPMOb optimizations over a broader range
of $\ththr$ and $\Nadj$ is summarized in Fig.~\ref{fig:fb_nm_rc_gpmob}.
For each optimization, final values of the field error objective $f_B$ and
magnet quantity $N_\text{mag}$ obtained by the GPMOb algorithm are shown 
relative to the corresponding values of the RC solution to which each
optimization was initialized. 
Overall, the solutions broadly exhibited a trade-off between the objectives
of high field accuracy and low magnet quantity. In general, as $\Nadj$ 
increased, $f_B$ increased and magnet quantity decreased. 
For a number of combinations of $\ththr$ and $\Nadj$ tested, the GPMOb algorithm
was able to improve the RC solution by reducing both field error and 
magnet quantity. 

While reductions in $f_B$ could be sustantial (about 40\% in
some cases), the reduction in magnet quantity was less significant
(generally 2-9\% in cases that also had improved $f_B$). In particular, the  
magnet quantities in these solutions far exceeded those of some GPMOb solutions 
initialized with empty magnet arrays: for example, the GPMOb$_{45^\circ,26}$
solution initialized with an empty array had 24,858 magnets per half-period,
or 32\% fewer magnets than in the RC solution. Nevertheless, these results
indicate a potential for the GPMOb algorithm to improve and fine-tune 
solutions produced by the RC algorithm.

\subsection{GPMOb-RC-GPMOb}
\label{sec:gpmob-rc-gpmob}

We performed another series of composite optimizations to determine whether
the RC approach could improve the output of a GPMOb optimization. As a test
case for a GPMOb solution to be improved, we used the output of the
GPMOb$_{45^\circ,26}$ optimization initialized to an empty magnet array
as described in Sec.~\ref{sec:gpmob_results}. This solution was used as the
initial guess for the RC algorithm. In addition, the RC algorithm was 
restricted to optimize the dipole moments only in 24,858 positions per 
half-period where magnets had been placed in the GPMOb$_{45^\circ,26}$ 
solution. This is in 
contrast to the RC solution with no GPMOb initialization described in 
Sec.~\ref{sec:rc_summary}, for which the RC algorithm was free to place magnets 
in any or all of the 58,258 positions per half period available in the input
arrangement. Finally, the output of this RC procedure was further refined
with a second GPMOb optimization; hence the designation ``GPMOb-RC-GPMOb''
for the composite approach described here. When referring to the solution 
output at the second stage, we will use the designation ``GPMOb-RC''.

\begin{figure}
  \begin{center}
  \includegraphics[width=0.48\textwidth]{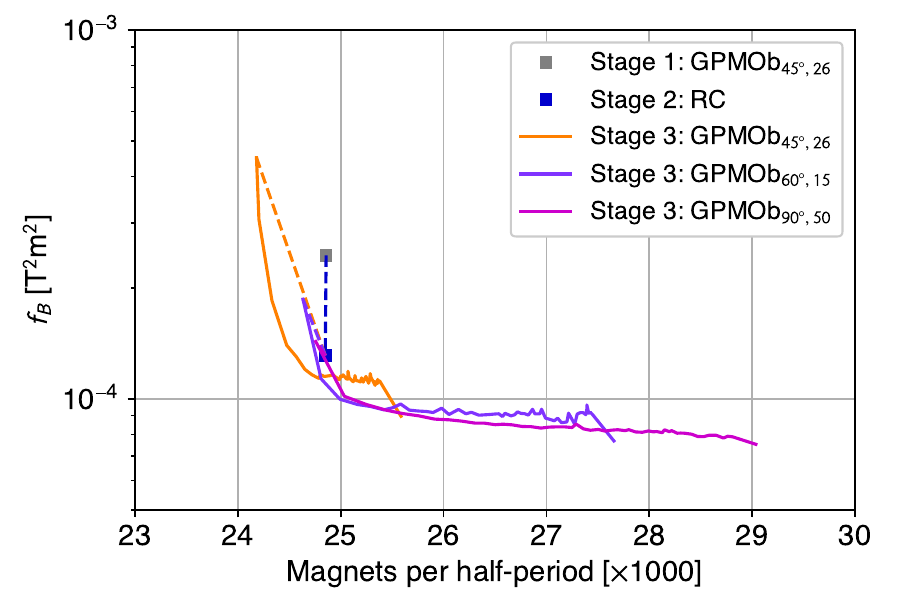}
  \caption{Iteration data from magnet optimizations using the three-stage
           GPMOb-RC-GPMOb approach. The output (final iteration) of the 
           first stage (GPMOb) is shown as the gray square. The output of the 
           second stage (RC) is the blue square. Iteration data for three
           different third-stage (GPMOb) optimizations are shown as solid
           curves with dashed lines indicating the changes in $f_B$ and
           quantity occuring after the initial backtracking procedure.}
  \label{fig:iterations_grg}
  \end{center}
\end{figure}

Iteration data from some sample GPMOb-RC-GPMOb optimizations are shown in
Fig.~\ref{fig:iterations_grg}. For the first stage (GPMOb$_{45^\circ,26}$) 
optimization, only the final interation is shown, as the gray square. 
In the second stage, an RC optimization was applied to the magnets placed
in the first-stage solution. This resulted in a decrease in $f_B$ and a slight
decrease in quantity, as shown by the blue square. For the third stage,
multiple GPMOb optimizations were performed with different $\ththr$ and $\Nadj$.
Iteration data from three of these are shown as the solid curves. In each
case, the jump in $f_B$ and corresponding drop in quantity following the first
backtracking procedure is indicated with
a dashed line of corresponding color. As was seen in 
Fig.~\ref{fig:iterations_rc_gpmob}, the final iteration data points from the 
(Stage 3) GPMOb optimizations appear to exhibit a sudden drop in $f_B$ and
increase in quantity; this is again because all points except the last one
are taken directly after backtracking is performed. 

Fig.~\ref{fig:iterations_grg} illustrates that both the second and third 
stages of the combined GPMOb-RC-GPMOb approach can improve upon the output of
the previous stages by decreasing both field error and magnet quantity.
For the third-stage (GPMOb) optimizations plotted, solutions with improvements 
in both metrics can be obtained from the earlier iterations; whereas later
iterations yield solutions with higher field accuracy at the expense of 
more magnets. Note, however, that all solutions shown on the plot have 
substantially fewer magnets than solutions produced by the RC and 
RC-GPMOb approaches. In addition, many of the solutions obtained in the
third stage have comparable or lower $f_B$ than those of the RC
and RC-GPMOb solutions.

\begin{figure}
  \begin{center}
  \includegraphics[width=0.48\textwidth]{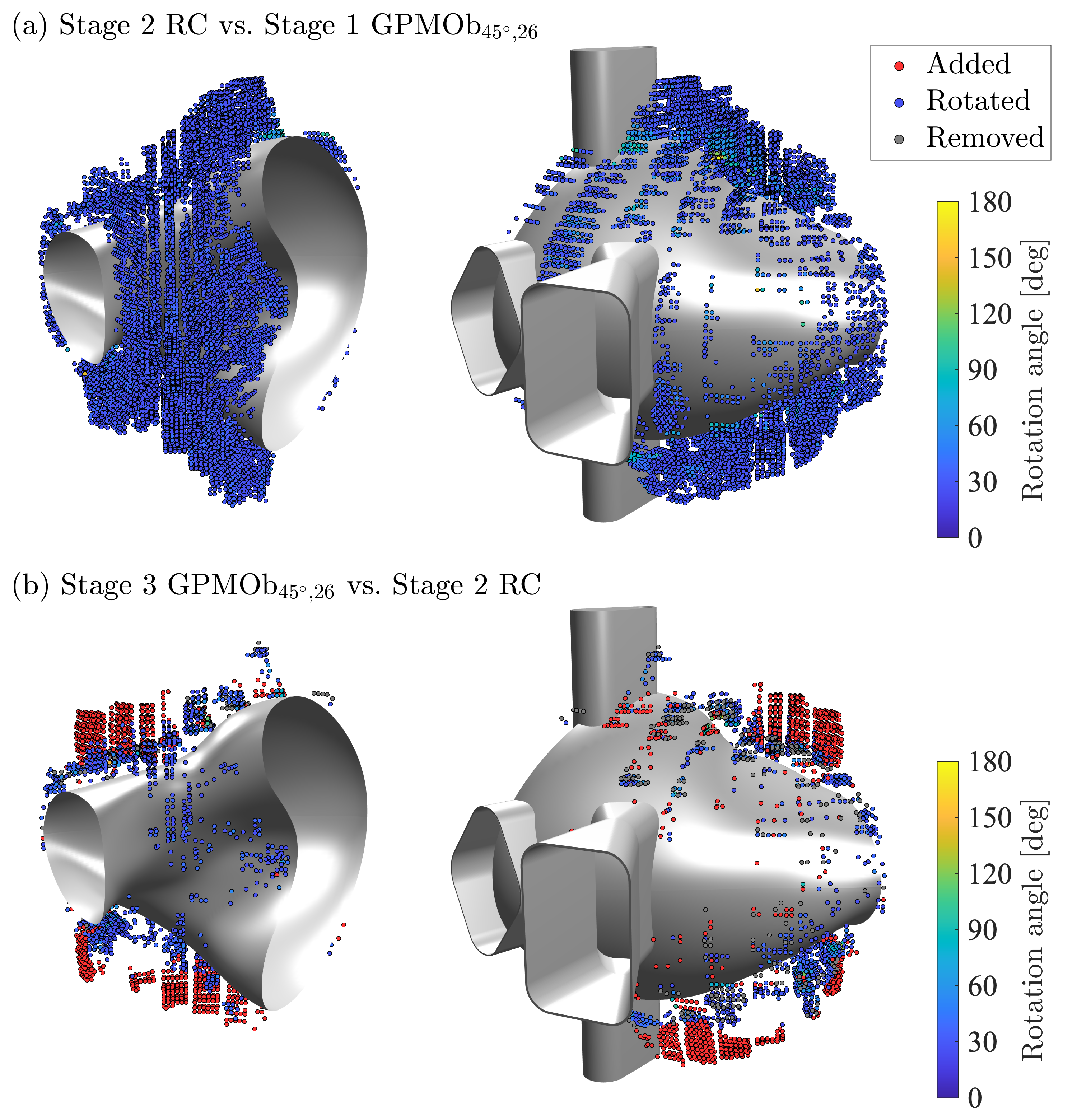}
  \caption{Depictions of the changes made to each dipole moment by the second
           (RC) and third (GPMOb) stages of an example GPMOb-RC-GPMOb 
           optimization relative to their respective foregoing solutions.
           Magnets added in a stage in positions that were previously empty
           are shown as red circles at the locations of the magnet centroids.
           Magnets removed in a stage that had existed previously are shown
           as gray circles. Magnets whose dipole moment changes direction
           during a stage are shown with colors according to the angular
           difference. Magnets that remain the same during a stage are not 
           shown.
           (a) Solution of the second-stage RC solution relative to the
               first-stage GPMOb$_{45^\circ,26}$ solution;
           (b) Solution of the third-stage GPMOb$_{45^\circ,26}$ solution
               relative to the second-stage RC solution.}
  \label{fig:vdiff_grg}
  \end{center}
\end{figure}

For further insight on the impact of the different optimization stages on the
solution, Fig.~\ref{fig:vdiff_grg} illustrates the changes that the solution
undergoes on a magnet-by-magnet basis. Fig.~\ref{fig:vdiff_grg}a compares the 
second-stage RC solution to the first-stage GPMOb$_{45^\circ,26}$ solution;
Fig.~\ref{fig:vdiff_grg}b compares the third-stage GPMOb$_{45^\circ,26}$
solution to the second-stage RC solution. In each plot, magnets that undergo
changes from one stage to the next are indicated with circles at their 
respective centroids. Magnets that are added to a previously empty position
are shown as red circles. Empty positions that had previously been filled are
shown as gray circles. Magnets whose dipole moment points in a different
direction from the previous solution are shown as circles colored according
to the angle subtended between the former and latter dipole moment vectors.

The differences in the nature of the changes made at Stages 2 and 3 shown in
Fig.~\ref{fig:vdiff_grg} help to illlustrate the different effects of the
RC and GPMOb algorithms. Following the RC optimization in stage 2, 53\% of the
magnets from the first-stage GPMOb solution are rotated, in most cases by angles
between $25^\circ$ and $30^\circ$. By contrast, following the third-stage GPMOb 
optimization, only 6\% of the magnets from the second-stage solution undergo
rotations; the vast majority remain the same. A higher proportion of the 
rotated magnets underwent rotations by larger angles ($35^\circ$ to $65^\circ$).
In addition, the changes made in
stage 2 are distributed more or less evenly around the magnet array, 
whereas the changes made in stage 3 are concentrated near the top and bottom.
Finally, in stage 2 the quantity does not change much: 8 magnets per 
half-period are removed and, by constraint, none are added. In stage 3, 
292 magnets are removed and 1,438 are added per half-period, primarily 
at the top and bottom.

\begin{figure}
  \begin{center}
  \includegraphics[width=0.48\textwidth]{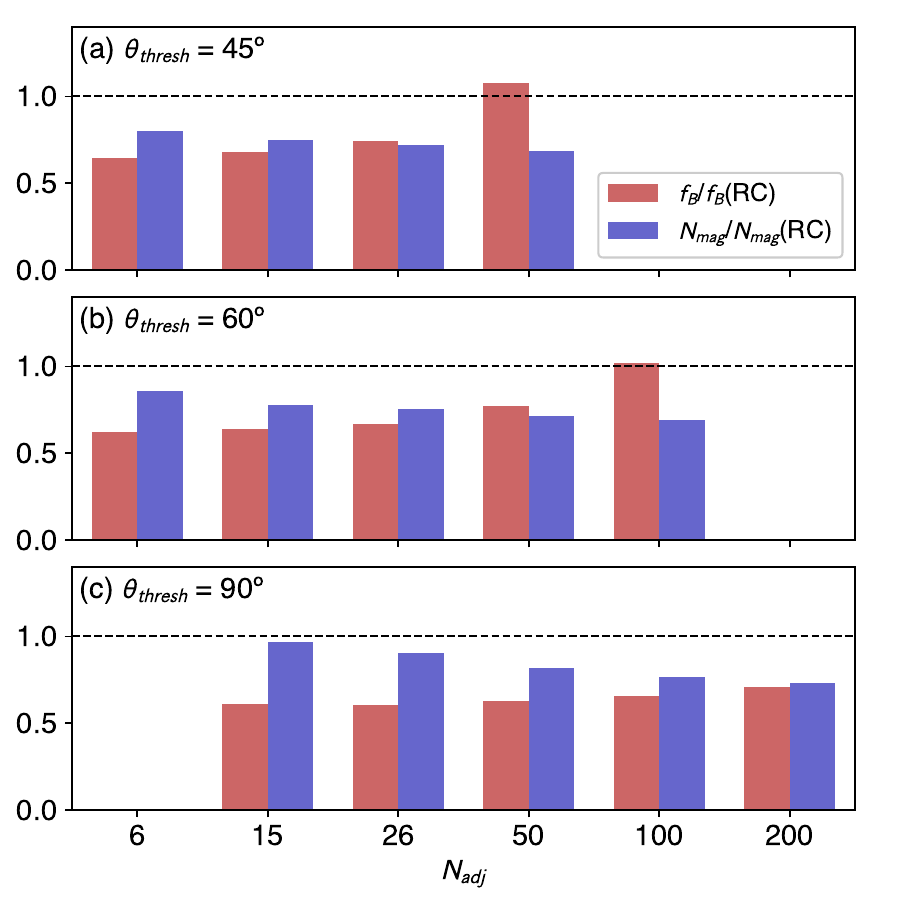}
  \caption{Performance of three-stage GPMOb-RC-GPMOb optimizations with 
           different $\ththr$ and $\Nadj$ employed during the third stage. 
           or each GPMOb solution, values of the error
           field objective $f_B$ and magnet quantity $N_\text{mag}$ are shown
           relative to the corresponding values for the RC solution described
           in Sec.~\ref{sec:rc_summary}.:
           $f_B(RC) = 1.20\times10^{-4}$~T$^2$m$^2$, 
           $N_\text{mag}(RC)=$~35,436 magnets/half-period.}
  \label{fig:fb_nm_grg}
  \end{center}
\end{figure}

Solutions from the three-stage approach are compared directly to the RC 
approach (Sec.~\ref{sec:rc_summary}) in Fig.~\ref{fig:fb_nm_grg}. For the
range of $\ththr$ and $\Nadj$ tested for the third stage of the 
GPMOb-RC-GPMOb optimization, values of $f_B$ and magnet quantity 
$N_\text{mag}$ from the final iteration of the stage-3 GPMOb optimization are 
shown relative to the corresponding values from the RC solution.
Several of these solutions exhibit substantial ($>$25\%) reductions in both
$f_B$ and $N_\text{mag}$ relative to the RC solution. This is in contrast to 
the results from the two-stage RC-GPMOb approach, for which solutions tended
to have substantially improved $f_B$ but only marginally reduced $N_\text{mag}$
(Fig.~\ref{fig:fb_nm_rc_gpmob}).

\begin{figure}
  \begin{center}
  \includegraphics[width=0.49\textwidth]{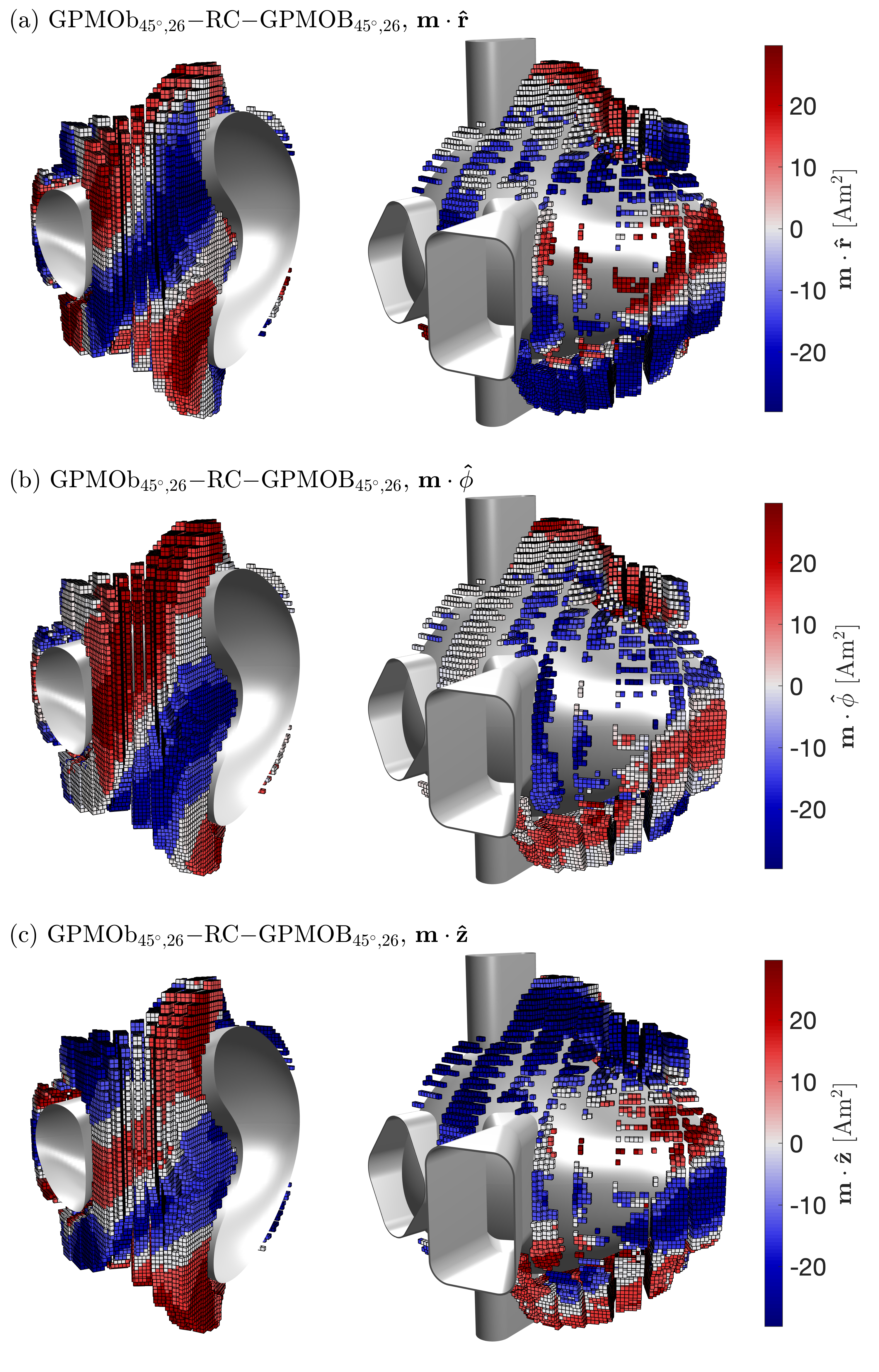}
  \caption{Renderings of the magnets in the solution produced by the 
           three-stage GPMOb-RC-GPMOb approach. Both the first and the third
           stages used backtracking parameters $\ththr=45^\circ$ and 
           $\Nadj=26$.
           (a) radial $r$ component, 
           (b) toroidal $\phi$ component,
           (c) vertical $z$ component.}
  \label{fig:vectors_grg}
  \end{center}
\end{figure}

Renderings of the GPMOb-RC-GPMOb solution in which the first and third stage 
used backtracking parameters $\ththr=45^\circ$ and $\Nadj=26$ is shown in
Fig.~\ref{fig:vectors_grg}. Overall, the distributions of each vector component
are qualitatively similar to those of the single-stage RC 
(Fig.~\ref{fig:vectors_rc}) and the GPMOb$_{45^\circ,26}$ 
(Fig.~\ref{fig:vectors_gpmob_45_26}) solutions. All three solutions utilize 
nearly all of the available magnet positions on the inboard side, with small 
differences the distribution patterns of the dipole moment vector across the
magnet array. On the outboard side, the GPMOb-RC-GPMOb solution has much
fewer magnets than the RC solution, particularly near the top and bottom
of the array. Relative to the single-stage GPMOb$_{45^\circ,26}$ solution, the 
placement of the magnets is very similar but there are slight differences in 
the dipole moment of each magnet, as on the inboard side. However, the 
dipole moments of the three-stage GPOMb-RC-GPMOb solution appear to be 
substantially better for field accuracy, yielding a reduction of 63\% in $f_B$
relative to the single-stage GPMOb solution. 

The three-stage solution shown in Fig.~\ref{fig:vectors_grg} represents the 
final GPMOb iteration with $\ththr=45^\circ$ and $\Nadj=26$ plotted in 
Fig.~\ref{fig:iterations_grg}. With $f_B=8.98\times10^{-5}$~T$^2$m$^2$ and
25,586 magnets per field period, this solution 25\% lower $f_B$ and 28\% fewer 
magnets than the original RC solution (Sec.~\ref{sec:rc_summary}). For an 
alternative solution that prioritizes further reduction in magnet quantity over 
increased field accuracy, one might choose, for example, an earlier iteration 
of the third-stage optimization with $f_B=1.16\times10^{-4}$~T$^2$m$^2$
and 24,721 magnets per field period. This would constitute a reduction of
3\% in $f_B$ and 30\% in magnet quantity relative to the original RC solution.

\section{Detailed comparison of solutions}
\label{sec:comparison}

In each of the optimization procedures studied in this paper, the scalar
objective function $f_B$, corresponding to the square integral of the 
normal component of the magnetic field on the target plasma boundary
(Eq.~\ref{eqn:fb}) was used as a proxy for the accuracy of the magnet
solution. Broadly speaking, lower values of $f_B$ correspond to more accurate
fields and better plasma confinement (more precisely, confinement 
characteristics that are closer to those of the target plasma). However,
as is well known in stellarator theory, the quality of plasma confinement for
a given $f_B$ can vary greatly depending on the spatial distribution of
the field errors. For example, error field distributions that resonate
with magnetic field lines on rational flux surfaces can be especially 
deleterious to confinement \cite{kerst1962a,rosenbluth1966a}. Hence, when 
comparing the merits of different solutions, it is worth performing more
detailed assessments of their plasma confinement properties.

To this end, we performed free-boundary equilibrium calculations for sample
magnet solutions obtained from each of the optimization approaches. The 
plasma equilibria confined by each magnet solution were determined with the
VMEC code operating in free-boundary mode \cite{hirshman1983a,hirshman1986a}.
The simulated plasma was assumed to have profiles of plasma current and 
pressure identical to those of the target plasma equilibrium, while the
external magnetic field was calculated from the magnet solution in
combination with the (fixed) toroidal-field coils. As a further assessment,
we used the NEO code \cite{nemov1999a} to evaluate profiles of the effective
ripple, a measure that correlates with neoclassical transport.

\begin{figure*}
    \begin{center}
    \includegraphics[width=\textwidth]{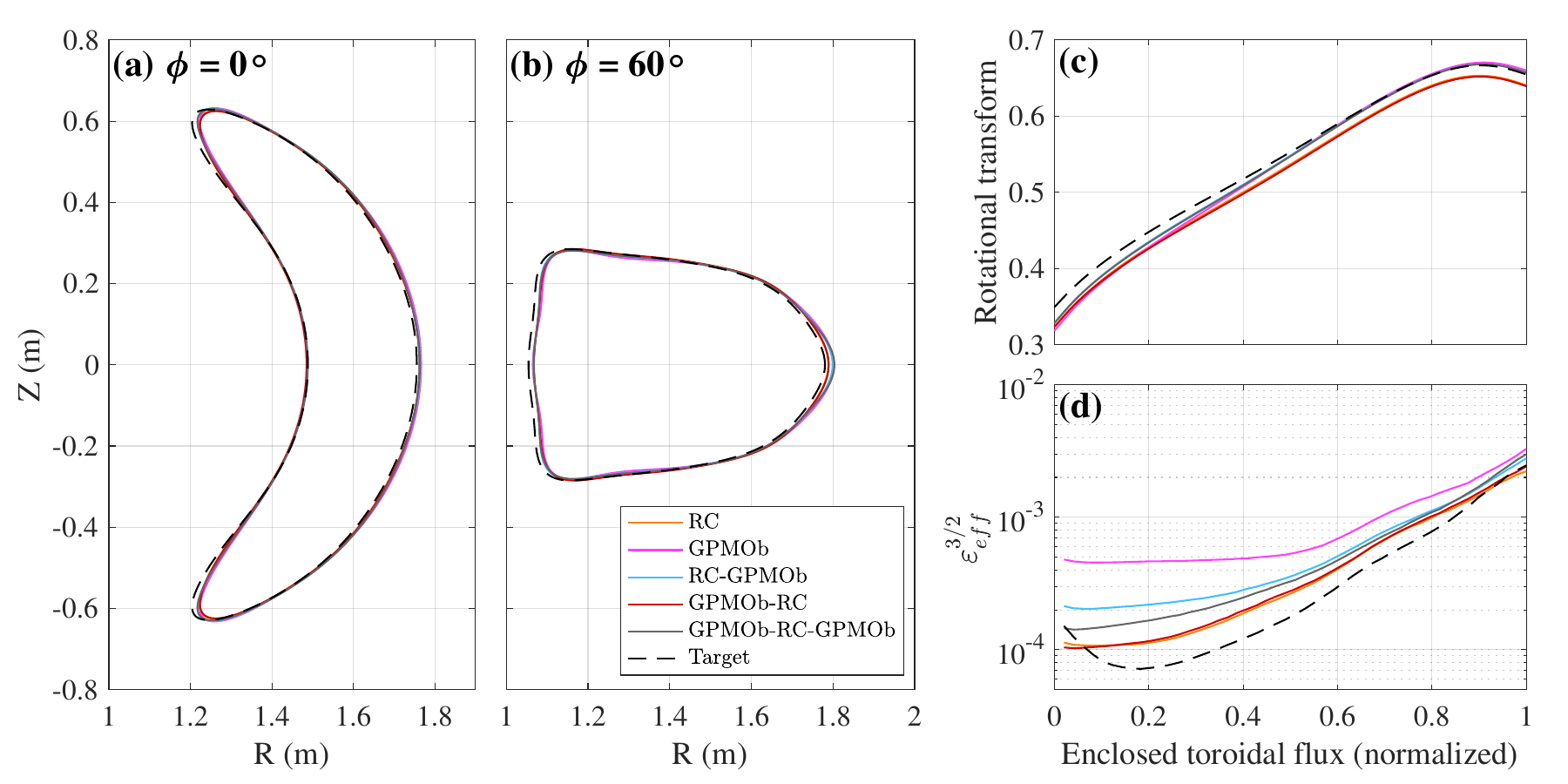}
    \caption{Characteristics of the plasma equilibria confined by the various
             magnet solutions obtained in the paper (solid colored curves) 
             relative to those of the target plasma equilibrium (black dashed
             curves).
             (a) Cross-sections of the plasma boundaries at toroidal angle
                 $\phi=0^\circ$;
             (b) Cross-sections of the plasma boundaries at toroidal angle
                 $\phi=60^\circ$;
             (c) Profiles of rotational transform; and
             (d) Profiles of effective ripple ($\epsilon^{3/2}$).}
    \label{fig:equil_comp}
    \end{center}
\end{figure*}

Results from these calculations are shown in Fig.~\ref{fig:equil_comp}.
Each subplot shows properties of plasma equilibria confined by various
magnet solutions (solid colored curves) compared to the corresponding
properties of the target plasma equilibrium (dashed black curves). These
properties include cross-sections of the plasma boundary at the two symmetry
planes (Fig.~\ref{fig:equil_comp}a-b), the profile of rotational transform
(Fig.~\ref{fig:equil_comp}c), and the profile of effective ripple
$\epsilon_\text{eff}^{3/2}$ (Fig.~\ref{fig:equil_comp}d). 

The magnet solutions under comparison include
the rounded continuous (RC) solution described in Sec.~\ref{sec:rc_summary},
the GPMOb$_{45^\circ,26}$ solution from Sec.~\ref{sec:gpmob_results},
the two-stage RC-GPMOb$_{45^\circ,26}$ solution from Sec.~\ref{sec:rc-gpmob},
and the solutions following the second stage (GPMOb$_{45^\circ,26}$-RC) and 
third stage (GPMOb$_{45^\circ,26}$-RC-GPMOb$_{45^\circ,26}$) of the
three-stage optimization from Sec.~\ref{sec:gpmob-rc-gpmob}. Since all 
GPMOb stages in this solution used the same values of $\ththr$ and $\Nadj$,
we will omit the subscripts for the rest of this section for brevity.

\begin{table}
    \begingroup
    \setlength{\tabcolsep}{6pt}
    \begin{tabular}{l c D{.}{.}{2.2}}
        \hline
                       & Magnets per & \multicolumn{1}{c}{$f_B$} \\
        Solution       & half period & \multicolumn{1}{c}{[$10^{-5}$ T$^2$m$^2$]} \\
        \hline
        RC             &      35,436 & 12.0 \\
        GPMOb          &      24,858 & 24.5 \\
        RC-GPMOb       &      32,836 & 8.55 \\
        GPMOb-RC       &      24,850 & 13.1 \\
        GPMOb-RC-GPMOb &      25,586 & 8.98 \\
        \hline
    \end{tabular}
    \endgroup
    \caption{Magnet quantities and $f_B$ values for the solutions compared
             in Sec.~\ref{sec:comparison}. All GPMOb optimizations shown
             here used $\ththr=45^\circ$ and $\Nadj=26$.}
    \label{tab:soln_qty_fb}
\end{table}

Overall, every magnet solution exhibits decent agreement with the target
plasma equilibrium. The most noticeable differences can be seen in the profiles
of effective ripple (Fig.~\ref{fig:equil_comp}d), although we note that the
values of $\epsilon_\text{eff}^{3/2}$ for all solutions are comparable to 
those of Wendelstein 7-X, for which 
$\epsilon_\text{eff}^{3/2} \geq 5\times10^{-4}$ \cite{beidler2021a}. 
Here, $\epsilon_\text{eff}^{3/2}$ for the single-stage GPMOb solution exceeds 
the target values by up to a factor of 6.4, whereas $\epsilon_\text{eff}^{3/2}$ 
for the RC and GPMOb-RC solutions both exceed the target values by a factor of 
no more than 1.6.

Notably, the offsets in effective ripple for each solution do not correlate
with the field error metric $f_B$. The $f_B$ values for each of these
solutions are shown in Table~\ref{tab:soln_qty_fb} along with the magnet
quantity. Perhaps not surprisingly, the solution with the highest field
error metric (GPMOb) also exhibited the greatest offset in effective ripple.
However, the solutions with the lowest $f_B$ (RC-GPMOb and GPMOb-RC-GPMOb)
did not exhibit the lowest offsets in ripple. Rather, the lowest offsets in
ripple are found for the solutions RC and GPMOb-RC, which have intermediate
values of $f_B$ among the solutions in Table~\ref{tab:soln_qty_fb}.
While these latter two solutions may not exhibit the lowest integral field
error, their lower $\epsilon_\text{eff}^{3/2}$ profiles provide evidence that 
the spatial distributions of their field errors are less deleterious for 
neoclassical confinement.

In light of these results, the GPMOb-RC solution may be the most advantageous
of the five under consideration in Table~\ref{tab:soln_qty_fb}. While it has 
a marginally higher $f_B$ than the original RC solution, it exhibits nearly
identical equilibrium and neoclassical transport properties while using 
30\% fewer magnets.

\section{Summary and conclusions}
\label{sec:conclusions}

In this paper, we have compared two recently-developed algorithms for 
optimizing permanent magnets for stellarator plasmas and demonstrated that
multi-stage optimizations utilizing the two algorithms in succession can 
produce substantially better solutions with that achieve better field accuracy
with fewer magnets. When used in isolation, the Rounded Continuous (RC)
approach achieved solutions with higher field accuracy but more magnets for
a given plasma and solution space,
whereas the Greedy Permanent Magnet Optimization algorithm with backtracking
(GPMOb) tended to find solutions with lower accuracy but fewer magnets.
Using the GPMOb optimization to refine the RC solution (i.e. the two-stage 
RC-GPMOb approach) attained solutions with both improved accuracy and slightly
fewer magnets than the RC solution with suitable choices of the backtracking 
parameters $\ththr$ and $\Nadj$. An alternate two-stage approach, GPMOb-RC,
wherein the second-stage RC optimization was restricted to optimize only the 
magnets that were placed in the first-stage GPMOb optimization,
attained a solution with nearly identical free-boundary equilibrium properties
to the original RC solution but with 30\% fewer magnets. Finally, applying
a third GPMOb stage to the GPMOb-RC solution attained solutions that
found solutions that reduced $f_B$ by 25\% or more relative to the RC solution,
although in at least one case, the neoclassical $\epsilon_\text{eff}^{3/2}$ 
metric was not as good as that of the GPMOb-RC solution.

These studies were enabled by some enhancements implemented in the GPMOb
algorithm within \textsc{Simsopt}. First, the user may now specify a list of
arbitrary allowable polarization vectors for each possible magnet position 
around the plasma. This enabled the usage of arbitrary magnet arrangements
while restricting the number of unique types of magnets required to construct
the solution. In this paper, all
solutions utilized three types of cubic magnets, distinguished by 
polarization orientation as illustrated in Fig.~\ref{fig:types}.
Second, the user may set an arbitrary threshold angle $\ththr$, specifying
the maximum allowable angular discrepancy in polarization vector for a nearby
pair of magnets to not be considered conflicting. 
Setting $\ththr$ lower effectively requires
the solution to have lower spatial gradients in polarization.
Finally, it is now possible to initialize the GPMOb optimization to an arbitrary
solution, allowing for multi-stage optimizations incorporating GPMOb and 
other algorithms.

With the three allowable
magnet types, together permitting 54 different dipole moments at each magnet 
position, the parameters $\ththr$ and $\Nadj$ had a significant impact 
on the nature of the solution.  We found that,
with a suitable choice of $\Nadj$, the GPMOb algorithm achieved the greatest
field accuracy with $\ththr=45^\circ$ or $60^\circ$. In general, the optimal
$\ththr$ will depend on the arrangement of magnet positions and our choice of 
allowable dipole moment vectors.

Previous work with the GPMOb algorithm produced magnet solutions
for a similar NCSX-like target plasma equilibrium with lower magnet volumes
and values of the $f_B$ metric less than $10^{-5}~$T$^2$m$^2$
\cite{kaptanoglu2023a}. However, it is important to note that the previous work 
used a different solution space. For example, its arrangement 
of possible magnet positions included positions directly adjacent to 
the vacuum vessel and left no gaps between magnets for support 
structures. By contrast, the arrangement used for this study (and in 
\cite{hammond2022a}) used an arrangement that reserved a minimum of 5.7~cm of
space between the magnets and the vessel (a distance equal to 18\% of the 
plasma minor radius), and additionally left space 
between adjacent blocks of magnets for mounting structures. For this plasma 
equilibrium and vessel geometry, enforcing gap spacing for an array 
magnets of rare-Earth strength has been shown to reduce the 
attainable field accuracy \cite{hammond2020a}. It is therefore not surprising
that the attainable values of $f_B$ in this work are somewhat higher. 
Nonetheless, free-boundary equilibrium modeling indicates that the optimizations
in this paper produced solutions with good neoclassical plasma confinement.

The results here suggest that sequential application of discrete and 
continuous optimization procedures is a promising way to address discrete
optimization problems with many degrees of freedom. The GPMOb and RC
algorithms appear to offer distinct and complementary approaches to finding
solutions within these high-dimensional spaces. 
We posit that the GPMOb algorithm is effective at identifying the most 
important locations for magnet placement, while the RC algorithm can find
areas of the solution space that make better use of a given set of magnet
positions. 

One interesting aspect of the optimizations performed in this study is that the
GPMOb-RC and GPMOb-RC(-GPMOb) optimization approaches could find solutions with
far fewer magnets than the RC solution without putting any explicit penalty
on the magnet quantity. Rather, the magnet quantity could be restricted
indirectly with suitable choices for the backtracking parameters $\ththr$ and 
$\Nadj$ used during the GPMOb stages. This seems to offer an 
advantage over an explicit quantity objective, which would be a more 
conventional approach to limiting magnet quantity. If a quantity objective
function were added to $f_B$ during either the GPMOb or the RC procedure,
a reduction in quantity would come at the expense of a lower reduction in field
accuracy (higher $f_B$). However, the GPMOb-RC(-GPMOb) approaches were able to 
produce solutions with both significantly reduced quantity (relative to RC) 
with a nearly equal (GPMOb-RC), or lower (GPMOb-RC-GPMOb), value of $f_B$.

\section*{Acknowledgments}

The authors would like to thank M.~Landreman and N.~A.~Pablant for their 
support of this project, as well as E.~J.~Paul for the helpful discussions.
This work was supported by the U.S.~Department of Energy under contract number 
DE-AC02-09CH11466 and award number DE-FG02-93ER54197.
The U.S. Government 
retains a non-exclusive, paid-up, irrevocable, world-wide license to publish or 
reproduce the published form of this manuscript, or allow others to do so, for 
U.S. Government purposes.

\section*{Conflict of interest}

K.C.H.~is a co-author of a patent application (PCT/US2023/064044) filed by
Princeton University that incorporates some of the methodology used in this 
work.

\section*{Data and code availability}

The data presented in this paper will be made available at 
\url{https://doi.org/10.34770/haxf-3c93}.
The open-source \textsc{Simsopt} codebase, in which the GPMOb algorithm
is implemented, may be accessed at 
\url{https://github.com/hiddenSymmetries/simsopt}. 
The open-source \textsc{Stellopt} suite of codes, used for plasma equilibrium
calculations, may be accessed at 
\url{https://github.com/PrincetonUniversity/STELLOPT}.
Access to the \textsc{Famus}
code, used for the continuous optimization in the RC approach, as well
as the \textsc{Magpie} code, used for generating the arrangement of magnet
positions used in this study, may be granted upon request.

\appendix

\section{Dependence of GPMOb solutions on backtracking parameters}
\label{sec:btparams}

\begin{figure}
  \begin{center}
  \includegraphics[width=0.48\textwidth]{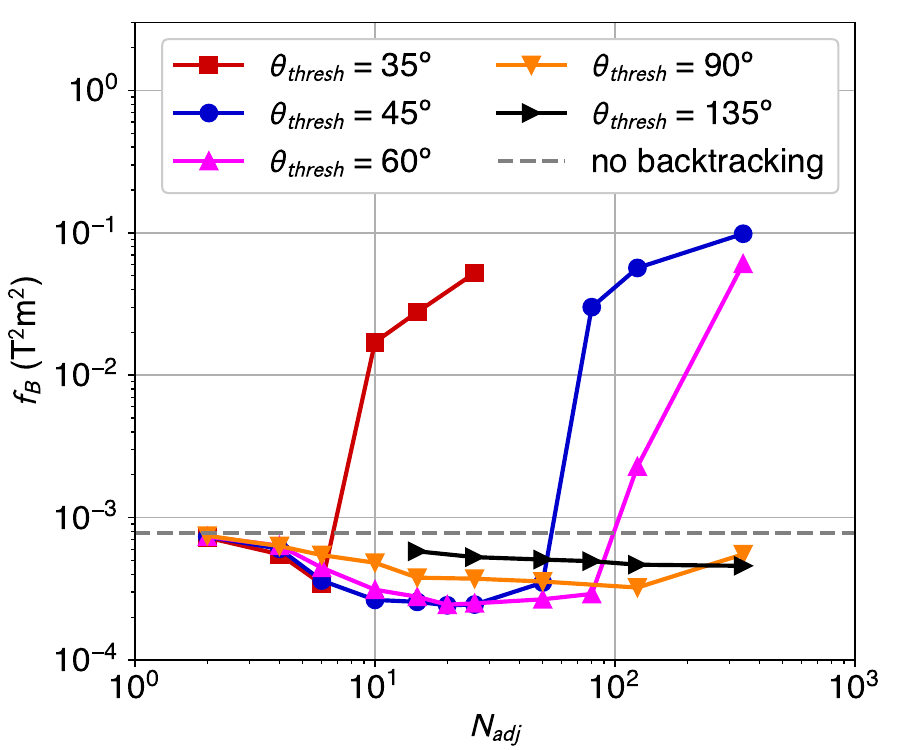}
  \caption{Values of the field error objective $f_B$ attainable with GPMOb
           optimizations with different values of $\ththr$ and
           $\Nadj$. The horizontal dashed line indicates the value obtained
           without backtracking. Solutions included magnets of the three types
           shown in Fig.~\ref{fig:types}.}
  \label{fig:fb_theta_nadj}
  \end{center}
\end{figure}

In Sec.~\ref{sec:gpmob_results}, it was observed that the backtracking 
parameters $\ththr$ and $\Nadj$ strongly influence the outcome of a GPMOb 
optimization. To identify the best parameters to use, we performed a set
of optimizations with a range of values of $\ththr$ and $\Nadj$. The results
are summarized in Fig.~\ref{fig:fb_theta_nadj}, which shows the values of $f_B$ 
achieved in each case.
The most accurate solutions were obtained with $\ththr$ values of
$45^\circ$ and $60^\circ$ and $\Nadj$ between 15 and 50. At these same
angles, the attainable $f_B$ increased sharply for $\Nadj > 50$, whereas
$f_B$ remained fairly consistent in this range of $\Nadj$ for 
$\ththr$ values of $90^\circ$ and $135^\circ$.

Some of the broader trends seen in Fig.~\ref{fig:fb_theta_nadj} can be 
understood by considering the limiting cases. For example, setting $\Nadj$ 
to zero would effectively shut off the backtracking procedure because that
would preclude any pairs of magnets from being checked for conflicting 
moments. It follows that, for any $\ththr$, the solution should approach
that of the case without backtracking as $\Nadj$ decreases toward zero.
The trends in Fig.~\ref{fig:fb_theta_nadj}, in which attainable $f_B$ values
for all $\ththr$ for $\Nadj<20$ approach the dashed line representing $f_B$ 
attained without backtracking, are consistent with this tendency.

Another limiting case would be obtained by setting $\ththr$ above the 
theoretical limit of $180^\circ$. This would effectively shut off the 
backtracking procedure because it would imply that no magnet pairs could be 
considered conflicting, irrespective of $\Nadj$. Hence, as $\ththr$ approaches
$180^\circ$, one expects the solution to approach that of the case without 
backtracking. The shift in $f_B$ as $\ththr$ is changed from $90^\circ$ to 
$135^\circ$ in Fig.~\ref{fig:fb_theta_nadj} appears to be consistent with this
trend, as the values become closer to the case without backtracking at most
values of $\Nadj$ tested.

We also note that setting overly stringent conditions on the backtracking
procedure can preclude accurate solutions. For example, increasing $\Nadj$
puts increasing restrictions on how much the distribution
of polarizations may vary spatially throughout the magnet array. However, some
degree of spatial variation among the magnets' dipole moments is necessary
to produce the spatial gradients in the magnetic field needed to confine
the plasma. Thus, for a given $\ththr$, it follows that there should be a
maximum feasible $\Nadj$ above which accurate solutions cannot be obtained.
As shown in Fig.~\ref{fig:fb_theta_nadj}, when $\ththr = 35^\circ$, this limit 
appears to be breached for $\Nadj \gtrsim 10$. When $\ththr = 45^\circ$, the
limit is breached for $\Nadj \gtrsim 60$. It follows that higher values of
$\ththr$ will tend to have higher limits for $\Nadj$, as increasing $\ththr$
permits greater spatial variation among neighboring magnets and therefore
moderates the restrictions on spatial variation imposed by increasing $\Nadj$.

%

\end{document}